\documentclass[10pt,letterpaper]{article}
\usepackage[top=0.85in,left=2.75in,footskip=0.75in,marginparwidth=2in]{geometry}

\usepackage[utf8]{inputenc}

\usepackage{cite}

\usepackage{nameref,hyperref}

\usepackage[right]{lineno}

\usepackage{microtype}
\DisableLigatures[f]{encoding = *, family = * }

\raggedright
\usepackage{changepage}

\usepackage[aboveskip=1pt,labelfont=bf,labelsep=period,singlelinecheck=off]{caption}

\makeatletter
\renewcommand{\@biblabel}[1]{\quad#1.}
\makeatother

\usepackage{lastpage,fancyhdr,graphicx}
\usepackage{epstopdf}
\fancyheadoffset[L]{2.25in}
\fancyfootoffset[L]{2.25in}

\usepackage{color}

\definecolor{Gray}{gray}{.25}

\usepackage{graphicx}

\usepackage{sidecap}

\usepackage{wrapfig}
\usepackage{tabularx}
    \newcolumntype{L}{>{\raggedright\arraybackslash}X}
\usepackage[pscoord]{eso-pic}
\usepackage[fulladjust]{marginnote}
\reversemarginpar

\begin{document}
\vspace*{0.35in}

% title goes here:
\begin{flushleft}
{\Large
\textbf\newline{Deep Generative Model Driven Protein Folding Simulations}
}
\newline
% authors go here:
\\
Heng Ma\textsuperscript{1},
Debsindhu Bhowmik\textsuperscript{1},
Hyungro Lee\textsuperscript{2},
Matteo Turilli\textsuperscript{2},
Michael T. Young\textsuperscript{1},
Shantenu Jha\textsuperscript{2,3},
Arvind Ramanathan\textsuperscript{4,*}
\\
\bigskip
\bf{1} Computational Sciences and Engineering Division, Oak Ridge National Laboratory, Oak Ridge, TN 37830
\\
\bf{2} RADICAL, ECE, Rutgers University, Piscataway, NJ 08854, USA
\\
\bf{3} Brookhaven National Laboratory, Upton, New York, 11973
\\
\bf{4} Data Science and Learning, Argonne National Laboratory, Lemont, IL 60439 
\\
\bigskip
* ramanathana@anl.gov

\end{flushleft}

\section*{Abstract}
Significant progress in computer hardware and software have enabled molecular dynamics (MD) simulations to model complex biological phenomena such as protein folding. However, enabling MD simulations to access biologically relevant timescales (e.g., beyond milliseconds) still remains challenging. These limitations include (1) quantifying which set of states have already been (sufficiently) sampled in an ensemble of MD runs, and (2) identifying  novel states from which simulations can be initiated to sample rare events (e.g., sampling folding events). With the recent success of deep learning and artificial intelligence techniques in analyzing large datasets, we posit that these techniques can also be used to adaptively guide MD simulations to model such complex biological phenomena. Leveraging our recently developed unsupervised deep learning technique to cluster protein folding trajectories into partially folded intermediates, we build an iterative workflow that enables our generative model to be coupled with all-atom MD simulations to fold small protein systems on emerging high performance computing platforms. We demonstrate our approach in folding Fs-peptide and the $\beta\beta\alpha$ (BBA) fold, FSD-EY. Our adaptive workflow enables us to achieve an overall root-mean squared deviation (RMSD) to the native state of 1.6$~\AA$ and 4.4~$\AA$ respectively for Fs-peptide and FSD-EY. We also highlight some emerging challenges in the context of designing scalable workflows when data intensive deep learning techniques are coupled to compute intensive MD simulations. 

% now start line numbers
% \linenumbers

% the * after section prevents numbering
\vspace{-0.2in}
\section{Introduction}
Multiscale molecular simulations are widely used to model
complex biological phenomena, such as protein folding,
protein-ligand (e.g., small molecule, ligand/ drug, protein)
interactions, and self-assembly~\cite{Dror_2012,Lee_2009}. However, much of these
phenomena occur at timescales that are fundamentally challenging
for molecular simulations to access, even with
advances in both hardware and software technologies~\cite{Bowman_2009}. Hence,
there is a need to develop scalable, adaptive simulation strategies
that can enable sampling of timescales relevant to these
biological phenomena.

Many adaptive sampling techniques~\cite{hruska2019extensible,Singhal_2009,Weber_2011,Doerr_2017,Mittal_2018,Shamzi_2018} have been proposed. All these techniques
share some similar characteristics, including (a) the need for efficient and
automated approaches to identify a small number of relevant conformational
coordinates (either through clustering and/or dimensionality reduction
techniques)~\cite{Shirts_2001,Savol_2011_ISMB,fox2019learning}, and (b) the identification of the ‘next’ set of simulations to
run such that more trajectories are successful in attaining a specific end
goal (e.g., protein that is well folded, protein bound to its target ligand,
etc.)~\cite{Mittal_2018,Shamzi_2018}. %There are numerous approaches to cluster simulations to characterize
%transition pathways from ensembles of bio-molecular simulations, such as
%Markov State Models and variational approach for molecular processes.

These adaptive simulations present methodological and infrastructral
challenges. Ref.~\cite{hruska2019extensible} provides important validation of
the power of adaptive methods over traditional ``vanilla'' molecular dynamics
(MD) simulations or ``ensemble'' simulations. Ref.~\cite{bala2019implementing}
highlights challenges of such workflows on high-performance computing
platforms.

% \textcolor{red}{Matteo/ Shantenu: Can you please add any of the material from
% the workflow side of things -- like the htbac or other systems that you have
% been working with where coupling of ML work loads exist with simulations?}

We recently developed a deep learning based approach that uses convolutions
and a variational autoencoder (CVAE) to cluster simulations in an unsupervised
manner~\cite{Bhowmik_2018}. We have shown that our CVAE can discover intermediate states from
protein folding pathways; further, the CVAE-learned latent dimensions cluster
conformations into biophysically relevant features such as number of native
contacts, or root mean squared deviation (RMSD) to the native state. 

We posit that the CVAE learned latent features can be used to drive adaptive
sampling within MD simulations, where the next set of simulations to run are
decided based on a measure of ‘novelty’ of the simulation/ trajectory frame
observed.

Integrating CVAE concurrently with large-scale ensemble simulations on
high-peformance computing platforms entails the aforementioned complexity of
adaptive workflows ~\cite{bala2019implementing}, while introducing additional
infrastructural challenges. These arise from the concurrent and adaptive
execution of heterogeneous simulations and learning workloads
requiring sophisticated workload and performance balancing, inter alia.

In this paper, we implement a baseline version of our deep learning driven
adaptive sampling workflow with multiple concurrent instances of MD
simulations and CVAEs. Our contributions can be summarized as follows:
\begin{itemize}
    \item We demonstrate that deep learning based approaches can be used to drive adaptive MD simulations at scale. We demonstrate our approach in folding small proteins, namely Fs-peptide and the $\beta$-$\beta$-$\alpha$-fold (BBA) protein and show that it is possible to fold them using deep learning driven adaptive sampling strategy.
    \item We highlight parallel computing challenges arising from the unique characteristics of the worklfow, viz., training of deep learning algorithms can take almost as much time as running simulations, necessitating novel developments to deal with heterogeneous task placement, resource management and scheduling.
\end{itemize}
Taken together, our approach demonstrates the feasibility of coupling deep learning (DL) and artificial intelligence (AI) workflows with conventional all-atom MD simulations. 
\vspace{-0.2in}
\section{Methods}
\vspace{-0.1in}
\subsection{Workflow description}
\vspace{-0.1in}
Two key components of the workflow include the MD simulation module and the deep-learning based CVAE module, which are described below. 
\paragraph{Molecular dynamics (MD) simulations:}
The MD simulations are performed on GPUs with
OpenMM 7.3.0~\cite{Eastman_2017}. Both the Fs-peptide and BBA systems were modeled using the Amberff99SB-ildn force field~\cite{Larsen_2011} in implicit Onufriev-Bashford-Case GBSA solvent model~\cite{Onufriev_2004}. The non-bonded interactions are cut off at 10.0 ~$\AA$ and no periodic boundary condition was applied. All the bonds to hydrogen are fixed to their equilibrium value and simulations were run using a 2 fs time step. Langevin integrator was used to maintain the system temperature at 300 K with a friction coefficient at 91 ps$^{-1}$. The initial configuration was optimized using L-BFGS local energy minimizer with tolerance of 10 kJ/mol and maximum of 100 iterations. The initial velocity is assigned to each atom from a Boltzmann distribution at 300 K. We also added a new reporter to calculate the contact matrix of $C_{\alpha}$ atoms in the protein (using a distance cut-off of 8 $\AA$ in hdf5 format using the MDAnalysis module~\cite{Michaud-Agrawal_2011,Beckstein_2016} that could be used as inputs to the deep learning module (described below). Each simulation run outputs a frame every 50 ps. 

\paragraph{Convolutional Variational Autoencoder (CVAE):}
Autoencoder is a deep neural network architecture that can
represent high dimensional data in a low dimensional latent space while retaining the key information~\cite{Doersch_2016}. With its unique hourglass shaped architecture, an autoencoder compresses input data into a latent space with reduced dimension and reconstructs it to the original data. We use the CVAE to cluster the conformations from our simulations in an unsupervised manner~\cite{Bhowmik_2018,Romero_2019}. Currently in our workflow, we use the number of latent dimensions as a hyperparameter (varying between ${3,4,5,6}$) and use the CVAE that most accurately reconstructs the input contact maps~\cite{Bhowmik_2018,Romero_2019}. CVAE was implemented using Keras/TensorFlow   and trained on a V100 GPU for 100 epochs. 

%Additionally, the variational layer constraints the data points to a normal distribution in latent space, in which way the latent embeddings will be evenly distributed and it links to any points in latent space to patterns in the original dataset. Convolutional layers are added before the feed forward layers, applying a filter to the input contact maps, which can improve the robustness of the network in recognizing the local patterns that represents local interactions between $C_{\alpha}$s from neighboring residues regardless of their positions.  

\begin{wrapfigure}{L}{7.5cm}
    \centering
    \includegraphics[width=7.5cm]{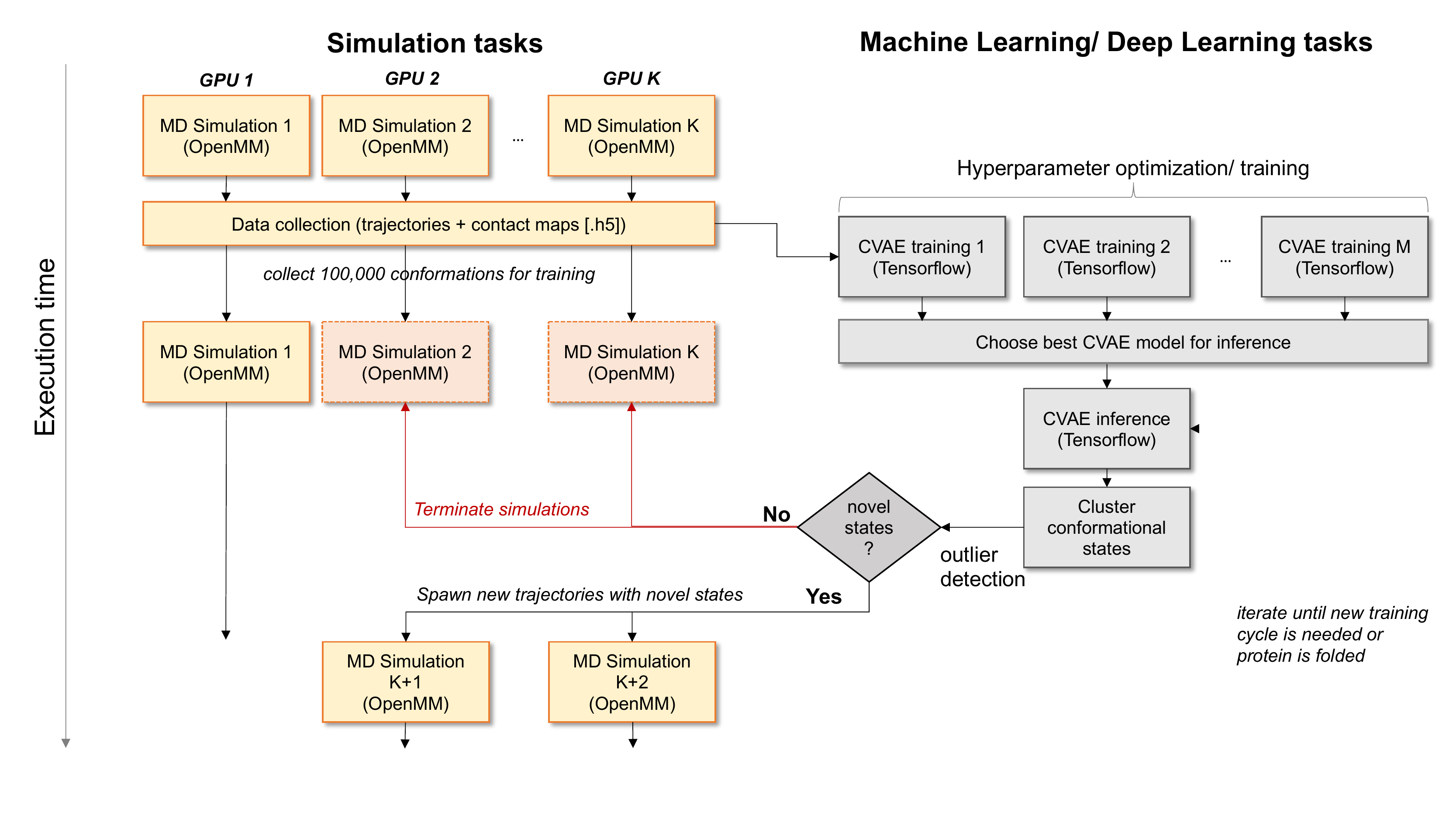}
    \caption{Deep generative model driven protein folding simulation workflow.}
    \label{fig:1}
\vspace{-0.2in}
\end{wrapfigure}

\paragraph{Assembling our workflow:} As illustrated in Figure ~\ref{fig:1}, our prototype workflow couples the two components. In the first stage, the objective is to initially train the CVAE to determine the optimal number of latent dimensions required to faithfully reconstruct the simulation data. We commence our runs as an ensemble of equilibrium MD simulations. Ensemble MD simulations are known to enable better sampling of the conformational landscape, and also can be run in an embarrassingly parallel manner. The simulation data is converted into a contact map representation (to overcome issues with rotation/translation within the simulation box) and are streamed at regular intervals into the CVAE module. The output from the first stage is an optimally learned latent representation of the simulation data, which organizes the landscape into clusters consisting of conformations with similar biophysical features (e.g., RMSD to the native state). Note that this is an emergent property of the clustering and the RMSD to the native state is not used as part of training data. 

In the second stage, our objective is to identify the most viable/ promising next set of starting states for propagating our MD simulations towards the folded state.  We switch the use of CVAE to infer from newly generated contact maps (from simulations) and observe how they are clustered. Based on their similarity to the native state (measured by the RMSD), a subset of these conformations are selected for propagating additional MD runs. The workflow is continued  until the protein is folded (i.e., conformations reach a user-defined RMSD value to the native state). 

\subsection{Implementation, Software and Compute Platform}

We used the Celery software to implement the aforementioned workflow. Celery is an asynchronous task scheduler with a flexible distributed system to process messages and manage operations, which enables real-time task processing and scheduling. The tasks can be executed and controlled by the Celery worker server asynchronously or synchronously. Celery applications use callables to represent the modules that are part of the workflow. Once called, the task client adds to the task queue a message where its unique name is referred so that the worker can locate the right function to execute. The flexibility of Celery framework enables real-time interfacing to manage resource and excise control over the task scheduling and execution. All tasks can be monitored and controlled directly by the object functions. By calling the tasks at different stages of their program, we simply build multi-task workflows, which supports a large volume of concurrent tasks with real-time interfacing. The use of Celery framework allows us to establish a baseline for estimating the compute requirements of our workflow. Our workflow comprises of two callables, namely that of MD simulations, and the CVAE used either in training or inference mode. 

We tested our deep learning driven adaptive simulation framework on the NVIDIA DGX-2 system at Oak Ridge National Laboratory (ORNL). The DGX-2 system provides more than 2 petaflops of computational power from a single node that leverages its 16 interconnected NVIDIA Tesla V100-SXM3-32GB GPUs. This enables us to distribute the MD simulations and CVAE training onto 12 and 4 GPUs respectively. All the components in the workflow are encapsulated within a Python script that manages the various tasks through Celery. It first initializes the Celery worker along with the selected broker, RabbitMQ. All 16 GPUs are then employed for MD simulations to first generate 100,000 conformers as the initial training data for CVAE. With 5 minute interval between iterations, the trained CVAEs continuously compress C$_{\alpha}$ contact map of conformers from MD trajectories into data points in latent space, which are subsequently evaluated with density-based spatial clustering of applications with noise (DBSCAN) for identifying outlier conformations~\cite{Ester_96}. We used DBSCAN for its relative simplicity and also to establish a baseline implementation of our code. For Fs-peptide, outliers were collected all four trained CVAE models and only CVAE with 6 dimensional latent space was applied for BBA outlier searching. In each iteration, the MD runs are examined for outliers. Simulations that pass an initial threshold of 20,000 frames (1 $\mu$s) for Fs-peptide and 10,000 (0.5 $\mu$s) for BBA, but do not produce any outliers for the last 5000 frames (250 ns of simulation time) are purged. With the available GPUs from such MD runs, new MD simulations are launched from the the outliers to ensure appropriate resource management and usage.

\vspace{-0.2in}
\section{Results}

In previous work~\cite{Bhowmik_2018}, we have shown that the CVAE can learn a latent space from the Fs-peptide simulations such that the conformations from the simulations cluster into distinct clusters consisting of folded and unfolded states. When parameters such as the RMSD (to the native sztate) and the fraction of native contacts are used to annotate the latent dimensions~\cite{Gsponer6719}, we showed that these latent representations correspond to reaction coordinates that describe how a protein may fold (beginning with the unfolded state ensemble). Thus, we posit that we can propagate the simulations along these low-dimensional representations and can drive simulations to sample folded states of the protein in a relatively short number of iterations.

\begin{figure}
    \centering
    \includegraphics[width=\textwidth]{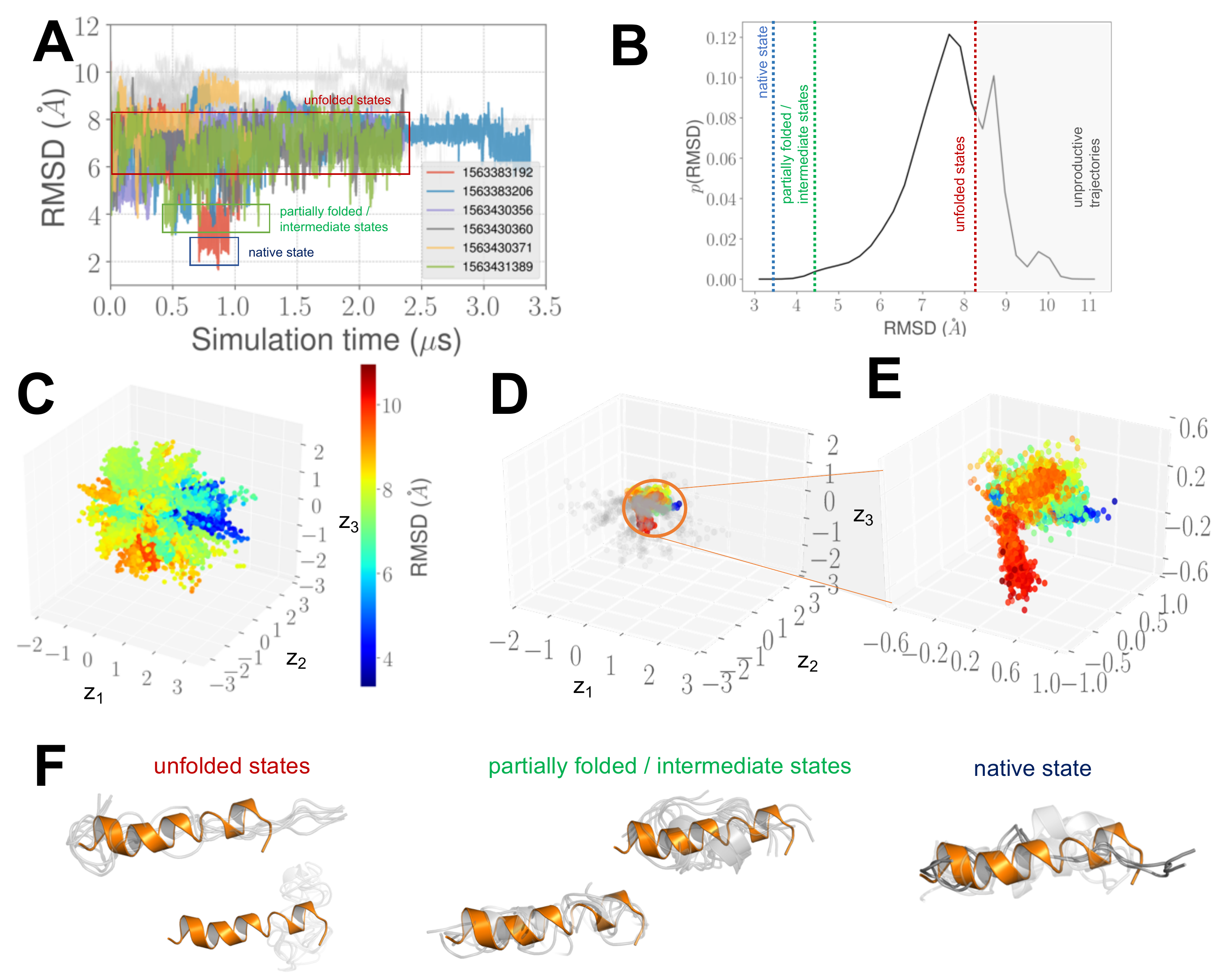}
    \caption{CVAE-driven folding simulations of Fs-peptide.(A) Root mean squared deviation (RMSD) with respect to the native/ folded state from the 31 trajectories generated using our adaptive workflow for the Fs-peptide system. Only productive simulations -- i.e., simulations that achieve a RMSD cut-off of 4.5 $\AA$ or less are highlighted for clarity. The rest of the simulations are shown in light gray. (B) A histogram of the RMSD values in panel (A) depicting the RMSD cut-off for identifying folded, partially folded, and unfolded ensembles from the data. The corresponding regions are also marked in panel (A). (C) Using the RMSD to the native state as a measure of foldedness of the system, we project the simulation data onto a three dimensional latent representation learned by the CVAE. Note that the folded states (low RMSD values highlighted in deeper shades of blue) are separated from the folding intermediate (shades of green and yellow) and the unfolded states (darker shades of red).(D)  A zoomed in projection of the last 0.5 $\mu$s of simulations generated along with the original projections (shown in pale gray, subsampled at every 100$^{th}$ snapshot). (E) highlights the same but just showing the samples from the last 0.5 $\mu$s to highlight the differences between folded and unfolded states. (F) shows representative snapshots from our simulations with respect to the unfolded, partial folded, and native state ensembles. Note that the cartoon representation shown in orange represents the native state (minimum RMSD of 1.6~$\AA$ to reference structure) determined from our simulations. }
    \label{fig:Fs-peptide-results}
\vspace{-0.1in}
\end{figure}

Figure \ref{fig:Fs-peptide-results} summarizes the results of our folding simulations of Fs-peptide. The peptide consists of 21 residues -- Ace-A$_5$(AAARA)$_3$A-NME -- where Ace and NME represent the N- and C-terminal end caps of the peptide respectively, and A represents the amino acid Alanine, where as R represents the amino acid Arginine. It is often used as a prototypical system for protein folding and adopts a fully helical structure as part of its native state ensemble~\cite{McGibbon2014}. Previous simulations used implicit solvent simulations using the GBSA-OBC potentials and the AMBER-FF99SB-ILDN force-field with an aggregate simulation time of 14 $\mu$s at 300K~\cite{McGibbon2014}. We used the same settings for our MD simulations and initiated our workflow. Summary statistics of the simulations are provided in Table \ref{tab:1}. A total of 90 iterations of the workflow was run to obtain a total sampling of 54.198 $\mu$s. Note that the sampling time of the MD simulations is an aggregate measure similar to the ones reported in previous studies. 

We began by examining the RMSD with respect to the native state from all of our simulations. As shown in Figure \ref{fig:Fs-peptide-results}A, 13 of the total of 31 simulations are unproductive -- i.e., they do not sample the native state consisting of the fully formed $\alpha$-helix. This is not entirely surprising given that the starting state consists of a nearly linear peptide with no residual secondary structures. Based on this observation, we posited that our CVAE model can be used to identify partially folded states from the simulations. We also examined the histogram of the RMSD values computed for each conformation with respect to the native state ensemble (Figure \ref{fig:Fs-peptide-results}B). Based on the histograms, we can reasonably choose a threshold of 3.1$\AA$ or less to depict the folded state ensemble, followed by 4.6~$\AA$ for partially folded states, and 8.3~$\AA$ for the unfolded states. Any trajectory that shows RMSD values beyond 8.3~$\AA$ are only sampling the unfolded state of the protein. 

\begin{table}[]
    \centering
    \begin{tabularx}{\linewidth}{c|LLLLL}
    \hline
         System & Total no. \newline simulations & Total \newline simulation time ($\mu$s) & (Shortest*, Longest) \newline simulations ($\mu$s) & Iterations & Min. RMSD ($\AA$) \\
    \hline
         Fs-peptide & 31 & 54.198 & 1.01, 3.4 & 90 & 1.6 \\
         BBA (FSD-EY)  & 45 & 18.562  & 0.517, 0.873& 100 & 4.44 \\
    \hline
    \end{tabularx}
    \caption{Summary statistics of simulations. *Only considering the simulations that pass the initial threshold. }
    \label{tab:1}
    \vspace{-0.2in}
\end{table}

The projections of all the 31 simulations onto the learned CVAE is depicted in Figure \ref{fig:Fs-peptide-results}C. Collectively, $z_1$-$z_3$ provide a description of the Fs-peptide folding process. Notably, much of the folded conformational states (highlighted in blue, indicating low RMSD to the native state) are clustered together. Similarly, the unfolded conformations (conformations colored in darker shades of red with higher RMSD to the native state ensemble) are also clustered together. Taking this further, we examined if the similarity in the conformations hold even with a smaller partition of the data (see Figures \ref{fig:Fs-peptide-results}D and E), namely the last 10\% of the overall simulation data. This can be treated as a test set from which new simulations are initiated. Notably, from these simulations we observe the presence of roughly three arms in the projections (Figure \ref{fig:Fs-peptide-results}E) consisting of: (1)  partially folded highlighted in shades of green/yellow, (2)  unfolded state ensemble highlighted in shades of red, and (3) a much smaller ensemble of folded states (highlighted in blue). 

For each of these states, we can also extract the structural characteristics with respect to the folded state (Figure \ref{fig:Fs-peptide-results}F). Many of the unfolded states do not consist of any secondary structural features (top and bottom left panels). The partially folded states consist of partial turns/ helical structures. The final folded state (with RMSD of 1.6~$\AA$) consists of most (if not all) helical turns in the protein. 
\vspace{-0.1in}
\subsection{Folding simulations of FSD-EY}
\begin{figure}
\includegraphics[width=\textwidth]{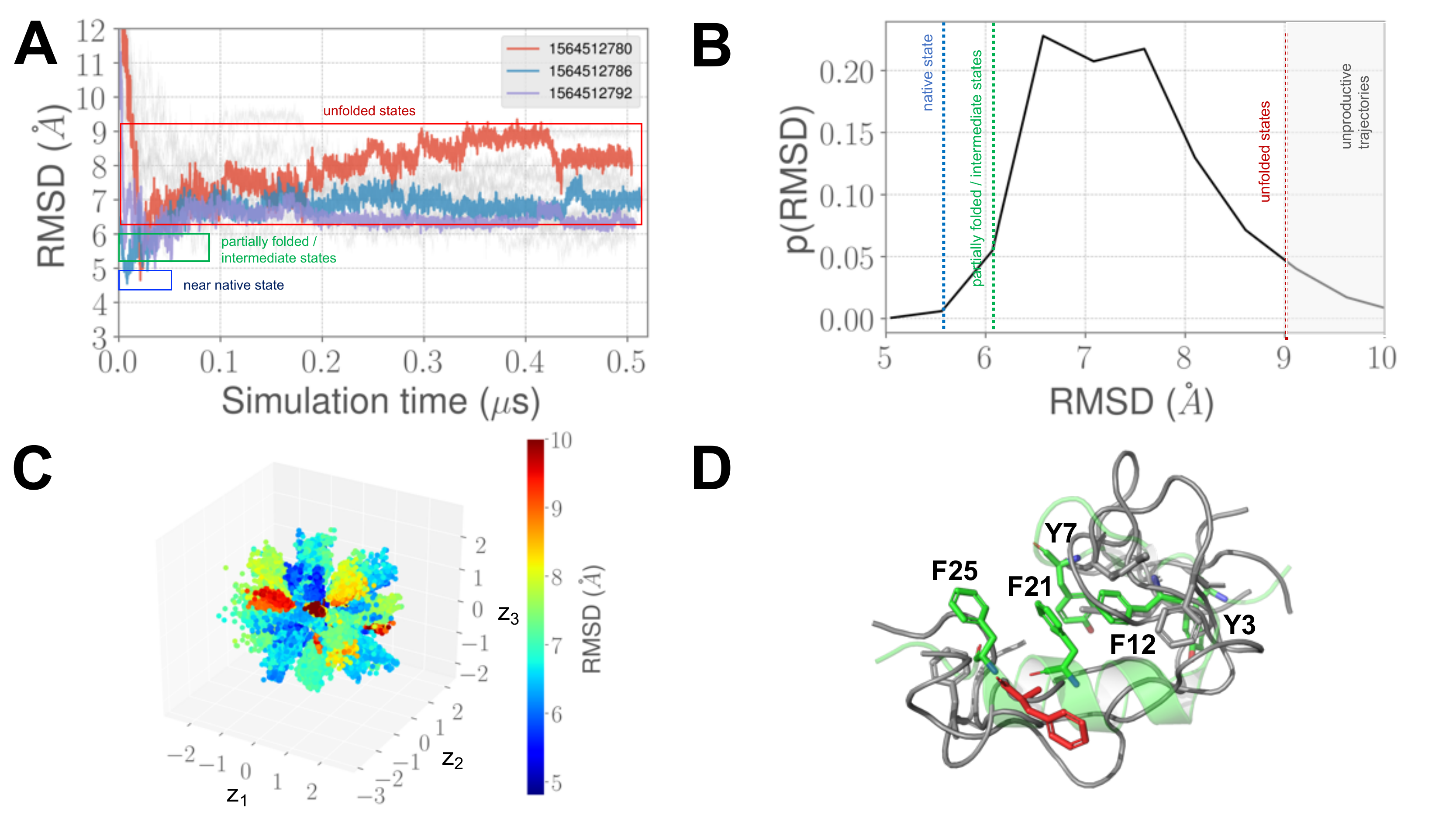}
\caption{CVAE-driven folding simulations of BBA-fold, FSD-EY. (A) RMSD plots with respect to the native state of FSD-EY depicting the near-native state (blue), partially folded states (green) and unfolded (red) trajectories similar to Figure \ref{fig:Fs-peptide-results}. (B) A histogram of the RSMD values to the native state. (C) The learned projections from the CVAE for the trajecotries; similar to the Fs-peptide system, we can observe the clustering of conformations based on their RMSD values to the native state. We have used a RMSD cut-off of 10$\AA$ to highlight states closer to the native state. (D) Although we could not fully fold the protein, we do observe the presence of a well-formed hydrophobic core except for one residue (F25) at the C-terminal end of the protein.}
\label{fig:BBAresults}
\vspace{-0.1in}
%\rule{3cm}{7cm}
\end{figure}

\vspace{-0.1in}
The BBA protein namely, FSD-EY is a designed protein that adopts a $\beta$-$\beta$-$\alpha$-fold in its native state; however this protein tends to be dynamic in solution~\cite{Lindorff-Larsen_2012,Sarisky_2011}. Similar to other zinc-finger proteins, the structure of the protein can potentially vary, and represents a challenging use-case for testing our workflow. As shown in Figure. \ref{fig:BBAresults}, our simulations do start with a completly unfolded state of the protein (average RMSD to native state is about 12~$\AA$. Using an aggregated MD sampling time of 18 $\mu$s, we note that we reach a RMSD value of 4.44~$\AA$. 

Although we do not sample the native state of the protein consisting of the $\beta$-$\beta$-$\alpha$-fold, we are still able to sample regions of the landscape that consist of a defined hydrophobic core consisting of the highlighted residues in Figure \ref{fig:BBAresults}D. Except for the dynamic C-terminal end, where the hydrophobic interactions between F21 and F25 are not entirely stable, the conformations that exhibit low RMSD values to the native state depict the presence of this hydrophobic core. We expect that extending these simulations further using the CVAE-driven protocol will enhance these interactions allowing it to fold completely. 
\vspace{-0.2in}
\section{Discussion}
As artificial intelligence (AI) and deep learning (DL) techniques become more pervasive for analyzing scientific datasets, there is an emerging need for supporting AI/DL coupled workflows to traditional HPC applications such as MD simulations. Our approach provides a proof-of-concept for how we can guide MD simulations to sample folded state ensemble of small proteins using DL techniques. The approach that we chose was based on building a generative model for protein conformations and identifying new starting conformations for additional MD sampling. Although the generative model was only used to identify novel conformations for extending our MD simulations, it nevertheless allowed us to guide the MD simulations towards sampling folded conformations of the protein systems we considered.

\begin{table}
    \centering
    \begin{tabularx}{\linewidth}{c|LLL|L}
    \hline
        System & DL training \newline (100 epochs; minutes) & Time per \newline epoch (seconds) & Inference time (ms/frame) & MD simulations \newline (ns per minute) \\
    \hline
        Fs-peptide & 7  &   5   & 5.13 & 1.25 \\
        BBA &   11  &   7   & 1.27 &  1.20 \\
    \hline
    \end{tabularx}
    \caption{Summary statistics of time taken by the individual components of our workflow: (1) train and infer from the CVAE for each system, and (2) running the MD simulation.}
    \label{tab:2}\vspace{-0.2in}
\end{table}

% Although DL approaches can take significantly longer time to train, we deliberately chose a simple, prototypic DL approach, namely CVAE, to train on our MD simulation data (Table \ref{tab:2}). As can be seen from the table, the computational throughput of training the CVAE model is on par with the throughput for running our MD simulations. \jhanote{Do you mean cost, or do you actually mean throughput (defined as some measure of load per unit time)?} This implies that in the time required to train our CVAE model, the MD simulation progresses only minimally -- meaning that interruptions to the MD simulation through start up of new simulations may not affect the workflow's overall performance. Also, our MD simulations were run using implicit solvent models, which significantly reduces computational time. Further, each contact map is no more than a couple of kilobytes of data and hence we did not require the utilization of intrinsic capabilities of the NVIDIA DGX-2, including the ability to potentially stream data across GPUs/ processors. 

Although DL approaches can take significantly longer time to train, we
deliberately chose a prototypic DL approach, namely CVAE, to train on our MD
simulation data (Table \ref{tab:2}). As can be seen from the table, the
computational cost of training and inference times for the CVAE model is
on par with the cost for running our MD simulations. %\jhanote{Do you
% mean cost, or do you actually mean throughput (defined as some measure of load
%per unit time)?} 
That is, within the time required to train our CVAE
model, our MD simulations progress only by about a nanosecond. Thus, starting up of new MD simulations based on the guidance received from our CVAE model will not affect the workflow's overall performance. Further, our MD 
simulations were run using implicit solvent
models, which also significantly reduces their computational times. 
Further, each contact map is no more than a couple of kilobytes of data and 
hence we did not require
the utilization of intrinsic capabilities of the NVIDIA DGX-2, including the
ability to potentially stream data across GPUs/ processors.
%\jhanote{Can we clarify the
% previous sentence please?} 

A primary motivation for this work was to use ML/DL based analysis to drive MD
simulations, and to calibrate results against non ML/DL driven approaches. In
Ref.~\cite{fox2019learning} Fox et al  introduced the concept of ``Effective
Performance'' that is achieved by combining learning with simulation and
without changing the traditional system characteristics.  Our selection of
physical systems, in particular the BBA peptide allows to provide a
coarse-grained estimate for the effective performance of CVAE based adaptive
sampling. Using Ref.~\cite{hruska2019extensible} as reference data, we find
that the effective performance of CVAE based sampling is at least a factor of
20 greater than "vanilla" MSM based sampling approaches. Our estimate is
based upon the convergence of simulated BBA structures to its reference
structures to within 4.5~\AA. Note that the ExTASY based simulations in
Ref.~\cite{hruska2019extensible} are at least two orders of magnitude more
efficient that reference DE Shaw simulations. In future work, we will extend
our effective performance estimate to Villin head piece (VHP) and refine our estimates
for BBA.

We expect that the concomitant increase in data sizes for larger MD simulations would necessitate the use of streaming approaches. Similarly, as the time required to train our DL models with data-/model- parallel approaches increases, it will require the use of emerging memory hierarchy architectures to facilitate efficient data handling/ transfer across compute nodes that are dedicated for training and simulation.  Further, data intensive techniques such as reinforcement learning and/or active learning could also have been used to guide our MD simulations. 

The requirements of the ML/DL driven simulations outlined in this paper are
representative of ML/DL driven adaptive workflows --- where the status of the
intermediate data analysis drive subsequent computations.  Adaptive workflows
pose significant challenges~\cite{bala2019implementing} compared to workflows
whose execution trajectory is predetermined a priori. Further, the integration
of diverse ML/DL approaches as the intermediate analysis driving subsequent
computations adds additional complexity, which includes but is not limited to
heterogeneous workload, load balancing and resource management. Scalable
execution and modern HPC platforms implies the need for specialized middleware
that support address these challenges. We are addressing these aspects as part
of ongoing work and future development built upon
RADICAL-Cybertools~\cite{bala2019implementing,turilli2019middleware}.

The source code, associated datasets including the generated simulations and deep learning models will be made available at the time of publication on http://www.github.com. 

% especially by integrating our approaches with the RADICAL-ENTK workflow
% system.

%\section{Conclusions}

\subsubsection*{Acknowledgements} We thank Vivek Balasubramanian and Jumana
Dakka for helpful discussions and early contributions. We also thank Chris Layton for helping set up our runs on the NVIDIA DGX-2 compute systems. We also acknowledge support by NSF DIBBS 1443054 and NSF RADICAL-Cybertools 1440677. This manuscript has been authored by UT-Battelle, LLC under Contract No. DE-AC05-00OR22725 with the U.S. Department of Energy. The United States Government retains and the publisher, by accepting the article for publication, acknowledges that the United States Government retains a non-exclusive, paid-up, irrevocable, world-wide license to publish or reproduce the published form of the manuscript, or allow others to do so, for United States Government purposes. The Department of Energy will provide public access to these results of federally sponsored research in accordance with the DOE Public Access Plan (http://energy.gov/downloads/doe-public-access-plan). This research used resources of the Oak Ridge Leadership Computing Facility at the Oak Ridge National Laboratory, which is supported by the Office of Science of the U.S. Department of Energy under Contract No. DE-AC05- 00OR22725.

\bibliography{Parco2019,hyperspace}

%This defines the bibliographies style. Search online for a list of available styles.
\bibliographystyle{abbrv}

\end{document}